\documentclass[preprint,12pt]{elsarticle}

\usepackage{graphicx}
\usepackage{xcolor}
\usepackage{changepage}
\usepackage{scrextend}
\usepackage{hyperref}
\usepackage{rotating}
\usepackage{amssymb}
\usepackage{lineno}

\journal{Physica A}

\begin{document}

\begin{frontmatter}

%\thanks{A footnote to the article title}%
\title{Lead-lag Relationships in Foreign Exchange Markets}% Force line breaks with \\
%druga verzija
%\title{Transformation equations for financial reference frames}
\author{Lasko Basnarkov$^{1,2}$}
\author{Viktor Stojkoski$^{2}$}
\author{Zoran Utkovski$^{3}$}
\author{Ljupco Kocarev$^{1,2}$}

\address{
$^{1}$SS. Cyril and Methodius University, Faculty of Computer Science and Engineering,  P.O. Box 393, 1000 Skopje, Macedonia}%
\address{
$^{2}$Macedonian Academy of Sciences and Arts, P.O. Box 428, 1000 Skopje, Macedonia}
\address{
$^{3}$Fraunhofer Heinrich Hertz Institute, Einsteinufer 37, 10587, Berlin, Germany}% 

\date{\today}% It is always \today, today,
             %  but any date may be explicitly specified

\begin{abstract}
%We study lagged correlations between one minute log returns on different rates from the foreign exchange market. Out of the statistically significant lagged correlations we construct a directed network. With the PageRank algorithm it was obtained that the most influential leaders in this network are certain market indexes quoted in corresponding domestic currencies. The respective lagged partial correlation coefficients have corroborated the findings from the ordinary correlations and show existence of mutual influence between other pairs. In a third approach, by calculation of linear regressions we have found that the leaders also Granger cause the laggers, thus confirming the direction of influence. These findings show that the market information does not spread instantaneously, contrary to the claims of the efficient market hypothesis.
Lead-lag relationships among assets represent a useful tool for analyzing high frequency  financial data. However, research on these
relationships predominantly focuses on correlation analyses for the dynamics
of stock prices, spots and futures on market indexes, whereas foreign exchange data have
been less explored. To provide a valuable insight on the nature of the lead-lag relationships in foreign exchange markets here we perform a detailed study for the one-minute log returns on exchange rates through three different approaches: i) lagged correlations, ii) lagged partial correlations and iii) Granger causality. In all studies, we find that even though for most pairs of exchange rates lagged effects are absent, there are many pairs which pass statistical significance tests. Out of the statistically significant relationships, we construct directed networks and investigate the influence of individual exchange rates through the PageRank algorithm. The algorithm, in general, ranks stock market indexes quoted in their respective currencies, as most influential. In contrast to the claims of the efficient market hypothesis, these findings suggest that all market information does not spread instantaneously.

%Altogether, these findings suggest that all market information does not spread instantaneously, contrary to the claims of the efficient market hypothesis.

%. From the statistically significant correlations we construct a directed network. With the PageRank algorithm it was obtained that the most influential leaders in this network are certain market indexes quoted in corresponding domestic currencies. The respective lagged partial correlation coefficients have corroborated the findings from the ordinary correlations and show existence of mutual influence between other pairs as well. In the third approach, by calculation of linear regressions we have found that the leaders also Granger cause the laggers, thus confirming the direction of influence. These findings show that all market information does not spread instantaneously, contrary to the claims of the efficient market hypothesis.

\end{abstract}


\end{frontmatter}

%\pacs{05.45.Tp,	%Time series analysis
%02.50.Sk, %Multivariate analysis
%87.23.Ge, %Dynamics of social systems
%02.50.-r  %Probability theory, stochastic processes, and statistics
%} 

%\keywords{Suggested keywords}%Use showkeys class option if keyword
                              %display desired
\maketitle

\section{Introduction}

Financial systems consist of many units influencing each other through interactions of different nature and scale, thereby exhibiting rather complex dynamics. The time series of the system observables, which capture these dynamics, are functions with pronounced random components. Hence, one requires convenient statistical tools to infer both the behavior of individual units and the overall performance of the financial system. In this aspect, a lot of effort has been put into developing tools for studying the pairwise relationships between assets as they provide direct estimate for the intensity of the mutual interactions. Besides this, the pairwise relationships also have particular practical relevance in construction of optimal portfolios and efficient asset allocations. As a result, various statistical approaches have been applied to plethora of financial markets, among which, equity prices at stock markets~\cite{plerou1999universal, mantegna1999hierarchical}, currencies at foreign exchanges~\cite{mizuno2006correlation, naylor2007topology} and even market indexes~\cite{eryiugit2009network}.

The majority of these studies mainly focus on relationships based on \textit{simultaneous} observations of the logarithmic returns of the examined assets. The results obtained from such studies are able to explain the mutual interactions only when the time needed for spreading of the news across the financial market is negligible in comparison to the period of calculation of log returns. The relevance of the results obtained in such situations is backed by the efficient market hypothesis. The hypothesis states that the current price of any asset traded in the market incorporates all relevant information and no  prediction regarding the future evolution would be profitable without taking additional risks~\cite{malkiel1970efficient}. Numerous studies confirm that in practice this usually holds when one considers log returns on daily or longer time scale. When shorter time scales are considered, such as one-minute log returns, one might expect significant delay effects. To address this issue, one usually resorts to the usage of \textit{lagged} approaches. These approaches give estimates for the relationship between the observations of the two quantities of interest which are apart from each other for certain period ~\cite{kawaller1987temporal,lo1990contrarian}. The variable whose values are delayed is called the \textit{lagger}, whereas the other one is the \textit{leader}.  %\textcolor{red} {In this work, we work only with lagged correlations, and sometimes for brevity, we omit emphasizing that, and use only the term correlations.}

%between two variables -- the \textit{leader} whose time series is taken as given, and the \textit{lagger}, whose time series is shifted at least one period~\cite{kawaller1987temporal,lo1990contrarian}.

Research on these lead-lag relationships predominantly focuses on correlation analyses of dynamics of stock prices or market indexes~\cite{huth2014high,xia2018emergence,valeyre2019emergence}\footnote{We note that the ordinary correlations are special case of the lagged ones, and are called zero-lagged or contemporary correlations in order to distinguish them from the latter.}. To the best of our knowledge, the examination of the lead-lag relationships in a foreign exchange market remains largely unexplored. To bridge this gap, here we study the relationships between one-minute log returns on exchange rates with a lag of one minute. We focus on returns lagged for one minute due to three reasons. First, even though past research suggests that the lead-lag relationships can be felt in the price of the lagger even up to 45 minutes later, as trading becomes more and more automatic, the lagging period has decreased dramatically to typical value measured in seconds~\cite{huth2014high}, or even milliseconds~\cite{dao2018ultra}. Second, the foreign exchange is known to be a very liquid one, with most of the transactions being ordered automatically and one can not expect significant lags that span longer than one minute. This reasoning was supported by studying lagged effects with two-minute lags. In this case, we found that although such lagged relationships might appear sporadically, the pattern is not persistent as it is when one-minute lags are considered.

We consider three different approaches to estimate the lead-lag relationships: i) lagged correlations, ii) lagged partial correlations and iii) Granger causality. The lagged correlations provide a direct estimate for the intensity of linear lead-lag relationship between two exchange rates. The lagged partial correlations extend this approach by eliminating the possible serial autocorrelation of the lagger and the contemporaneous correlation of both rates under study. Finally, the Granger causality~\cite{granger1969investigating} statistically tests for possible causal relationship between two exchange rates by considering the predictive potential by using past returns. We find that, with few exceptions, the lagged correlations between exchange rate pairs which relate currencies do not pass tests of statistical significance, which is in accordance with the efficient market hypothesis. However, there are many statistically significant lagged correlations, particularly when one considers rates which involve stock market indexes. In these correlation coefficients, the stock market indexes, which are known to have slower dynamics than the currency exchange raters, appear as leaders. Interestingly, this is opposite to previous findings which suggest that assets with faster dynamics behave as leaders \cite{kawaller1987temporal, lo1990contrarian, toth2006increasing}. The partial correlation analysis further confirmed our findings. The Granger causality analysis, in addition to providing a test for the findings in the correlation analysis, revealed that the leaders in the lead-lag relationships also increase the predictive ability for the determination of future returns of the lagger. Based on these observations one could believe that the information from the leader towards the lagger does not transfer simultaneously, but it can be seen as a process that stretches for a certain period. 

%In this work we study the correlations between one minute log returns on exchange rates with lag of one period, that is one minute. Foreign exchange market is very liquid one, with most of the transactions made automatically and one can not expect any lagging longer than this period. We have found that, with few exceptions, the lagged correlations between pair of exchange rates between currencies do not pass tests of statistical significance, which is in concordance with the efficient market hypothesis. However, there are many statistically significant lagged correlations when one considers rates which involve stock market indexes. Moreover, these rates appear as leaders, which is opposite to the previous findings, since these have slower dynamics than the exchange rates between two currencies. We have furthermore considered Granger causality analysis between pairs of exchange rates and found that the leaders in lead-lag relationships also increase the predictive ability for determination of future return of the follower. On base on these observations one could believe that the information from the leader towards the lagger does not transfer simultaneously, but it can be seen as a process that stretches for certain period. Moreover, we consider partial lagged correlations which are targeted to uncover the influence of the past log return of the leader to the current return on the lagger, when serial autocorrelation of the lagger and the contemporaneous correlation between them is excluded. The values of these correlations are in concordance to the other findings.

Out of the statistically significant pairwise correlations and causality relationships we obtained matrices, which can be regarded as directed networks that describe the effect of one asset towards another. To infer the most influential leaders, in the spirit of~\cite{brin1998anatomy}, we apply the PageRank algorithm. PageRank is a widely used procedure that has been applied in original or modified form to various domains \cite{ white2003algorithms, ma2008bringing, falagas2008comparison, radicchi2011best}. The resulting ranking allows us to infer the major sources of information spillover in the studied financial network.

%which is a tool for ranking nodes in various directed networks \textcolor{red}{Razni referenci}. This approach also puts at the top certain stock market indexes. These seem to be major sources of information in this financial network

%From the pairwise statistically significant correlations and causality relationships one obtains matrices, which can be regarded as directed networks because the influence they represent is asymmetrical. The resulting networks are much sparser than those corresponding to the contemporaneous correlations. We apply the Page Rank algorithm in order to determine the most influential leaders \cite{brin1998anatomy}. As expected, this approach puts at the top the stock market indexes.

The outline of the paper is as follows. In Section 2 we introduce the methods applied for the estimation of the lagged correlations, lagged partial correlations and causality, as well as the generation of the resulting networks. Section 3 presents the empirical results. The last section summarizes our findings and provides directions for future research.

\section{\label{sec:methods}Methods}

\subsection{Lagged correlations\label{Lagged_corr_theory}}

The dynamics of the prices of financial assets are known to be non-stationary. Hence, one can not simply use them to examine the relationship between different assets. Nevertheless, the logarithmic return,
\begin{equation}
    r_n = \log p(t_{n+1}) - \log p (t_{n}),
    \label{eq:log_ret_def}
\end{equation}
where the price $p$ is assumed to be observed at discrete moments $t_n$, is usually assumed to be weakly stationary~\cite{fan2017elements}. This property of log returns makes them more appropriate quantities for uncovering statistical relationships between assets.

Assuming that we have observations of the log returns $r_i$ and $r_j$ of two prices at the series of discrete moments $t_n$, the $\tau$-lagged covariance between them is given by
\begin{equation}
C_{i,j}(\tau) = \langle \left[r_i\left(t_n\right) -  \langle r_i \rangle \right] \left[r_j\left(t_n + \tau \right) -  \langle r_j \rangle \right] \rangle,
\label{eq:lag_cov_def}
\end{equation}
where the angular brackets denote averaging. We emphasize that one should keep the order of indices since in this notation the first index is the leader, while the second denotes the lagger. In general, the lagged covariances are not commutative $C_{i,j}(\tau) \neq C_{j,i}(\tau)$, in contrast to the ordinary, zero-lag covariances. Since the log return of one price might typically deviate more than that of the other one, a more appropriate quantity is the correlation coefficient
\begin{equation}
\rho_{i,j}(\tau) = \frac{C_{i,j}(\tau)}{\sqrt{C_{i,i}(0) C_{j,j}(0)}} = \frac{C_{i,j}(\tau)}{\sigma_{i} \sigma_{j}}.
\end{equation}
In the last expression the notation is simplified by using the standard deviation $\sigma_i = \sqrt{C_{i,i}}$.

%\textcolor{red}{In this work we determine the network of lagged correlation between assets by keeping only the correlations which exceed certain level of $p$-value, as is done in \cite{curme2015emergence}}.

\subsection{Lagged partial correlations \label{Lagged_partial_theory}}

In reality, the existence of lagged correlation can be due to one exchange rate being the driving force of the other, or due to the combination of mutual contemporaneous correlation and lagged autocorrelation. In order to isolate this potential effect, we calculate the partial correlation coefficient. For three random variables $X$, $Y$, and $Z$ with respective pairwise correlations $\rho(X,Y)$, $\rho(Y,Z)$, and $\rho(X,Z)$, the partial correlation between $X$ and $Y$ with $Z$ considered known, is defined as
\begin{equation}
    \rho(X,Y|Z) = \frac{\rho(X,Y) - \rho(X,Z)\rho(Y,Z)}{\sqrt{[1-\rho^2(X,Z)][1-\rho^2(Y,Z)]}}.
\end{equation}
Therefore, when the current return of the lagger $j$ has the role of known variable, the correlation between its next value and current return on the leader $i$ is
\begin{equation}
    \rho_{i,j}^P(\tau) = \frac{\rho_{i,j}(\tau) - \rho_{i,j}(0)\rho_{j,j}(\tau)}{\sqrt{[1-\rho_{i,j}^2(0)][1-\rho_{j,j}^2(\tau)]}}.
\end{equation}
%We point out that there might also be other exogenous variables driving both exchange rates that are not removed with this procedure. 
The last expression is useful when one has calculated already the values of the ordinary correlations and thus would immediately obtain the partial one. Another approach for calculation of the partial correlation is based on linear regression~\cite{baba2004partial}. It can be described as follows. Let $Z$ be the variable whose influence has to be removed, the two other variables being $X$ and $Y$, and their best estimates by observing $Z$ being  $\hat{X}_Z$ and $\hat{Y}_Z$ respectively. The respective partial correlation is then the ordinary correlation between the residuals $r_X = X-\hat{X}_Z$ and $r_Y = Y-\hat{Y}_Z$, given by
\begin{equation}
    \rho(X,Y|Z) = \rho(X-\hat{X}_Z, Y-\hat{Y}_Z).
\end{equation}
This approach allows for easier calculation of the statistical significance of the estimated partial correlation, and as such has been widely applied in studies of relationships between currencies and stocks \cite{kenett2010dominating, kenett2015partial, mai2018currency}.

%We mention that this approach of filtering out the influence of third asset on the correlation of two by partial correlation has already been applied in studies of relationships between currencies and stocks \cite{kenett2010dominating, kenett2015partial, mai2018currency}.

\subsection{Lagged correlation networks \label{Lagged_net_theory}}

%\textcolor{red}{Dali da napravime nenasochena mrezha kade shto kje gi stavime zbirot od apsolutni vrednosti kako vrska. Vaka imame jaki vrski koga megjusebno se pottiknuvaat. Pa posle da pravime zaednici ili nekoi grafovi. Ova bi se sporedilo so nasocheniot graf.}

As the lagged correlations are not commutative (in contrast to the contemporaneous correlations), the resulting network is directed. In order to keep the strength of influence we create a weighted network where the direction is from the lagger towards the leader with weight that is equal to the absolute value of the lagged correlation coefficient $w_{j,i} = |\rho_{j,i}|$. We consider absolute values since the sign of the correlation coefficient denotes only the direction of change of the exchange rate. Typically, extraction of the core of financial correlation networks is performed with the Minimal Spanning Tree (MST) sub-graph procedure~\cite{mantegna1999hierarchical} or the Planar Maximally Filtered Graph (PMFG)~\cite{kenett2010dominating, tumminello2005tool}. Because the obtained statistically validated lagged correlation networks are rather sparse, we may instead use the PageRank algorithm in order to determine the most influential leaders~\cite{brin1998anatomy}. The PageRank algorithm was originally used to rank web pages, by assuming that pages with more incoming links from others are more important. A particular feature which favors PageRank above other ranking procedures is that it intrinsically assigns higher weights to links that originate from important nodes. 

The application of PageRank first involves creating a row stochastic matrix from the weighted network with elements
\begin{equation}
    a_{j,i} = \frac{w_{j,i}}{\sum_k w_{j,k}}.
\end{equation}
The last matrix can be seen as a Markov chain transition matrix. In terms of Markov chain theory, the element $a_{j,i}$ corresponds to the probability of jumping from state $j$ to $i$. For such matrices arising from correlation networks, the bigger values within $j$-th row correspond to the columns $i$ associated with rates from which $j$ obtains larger impact. As such, the resulting rankings should provide information regarding the market news spreading among the the exchange rates included in the analysis. Higher ranking should correspond to exchange rates whose changes in returns will also have a higher overall impact in the overall system dynamics.

%Thus, the nodes with more incoming links, or links with more weight, have bigger importance in the network. %This is similar to the World Wide Web, where the pages with more incoming links from the other pages are considered as more important. An improved view on the directed graph of pages can be obtained to assigning more weights to links that originate in important nodes. This is the basic idea of the Page Rank algorithm that ranks the nodes according to their visiting frequency by a random surfer \cite{brin1998anatomy}.

\subsection{Granger causality \label{Granger_theory}}

%\textcolor{blue}{Ova e standrardna analiza koga imame vlijanie so zadocnuvanje (ili slichno kazhano) no od druga strane mozhe da bide i proverka za robustnost.}

%\textcolor{blue}{Citati \cite{granger1969investigating, lutkepohl2005new}}

The potential influence of the return on certain exchange rate on future returns on other exchange rates can be assessed by applying the causality analysis in the sense of Granger~\cite{granger1969investigating}. WIthin this framework, cause and effect relationship between a pair of two dynamical variables exists if knowledge of past values of one of them improves the prediction of the future of the other. %Granger introduced a formal tool for estimation whether such cause-and-effect interactions exist between two quantities. 
Granger causality is estimated as follows. Consider the linear regression of the nearest future value of some variable $x$ from its past $p$ terms
\begin{equation}
    x_{n+1} = \sum_{k=0}^{p-1} \alpha_k x_{n-k} + \varepsilon_n,
    \label{eq:self_regression}
\end{equation}
where $\varepsilon_n$ is the regression residual. The weights $\alpha_k$ are such that the residual variance $\sigma_{\varepsilon}^2$ is minimal, so the linear regression is optimal. Now, make an optimal combined regression by using $q$ past values of the variable $y$ as well
\begin{equation}
    x_{n+1} = \sum_{k=0}^{p-1} \alpha_k x_{n-l} + \sum_{l=0}^{q-1} \beta_l y_{n-l} + \eta_n.
\end{equation}
If some of the parameters $\beta_l$ are nonzero with certain statistical significance, which in turn leads to reduced variance $\sigma_{\eta}^2 < \sigma_{\varepsilon}^2$, then it is said that $y$ Granger causes $x$. If $x$ also Granger causes $y$ than one has a feedback system. The regression depth $q$ (and $p$ as well) depends on the problem at hand. We consider only one past step, as it has been observed that in foreign exchange markets the serial autocorrelation quickly vanishes.
Another reason for this choice is that, as it will be seen in the Data Section, due to the presence of gaps and periods of repeating identical values in the studied time series, there is a much smaller number of consecutive nonzero returns than the one required for a meaningful regression with more explanatory variables.
Therefore, our Granger causality setting for the effect of exchange rate $i$ on exchange rate $j$ is described as
%between different exchange rates, one could make a linear regression of the log return $r_j$ with its previous value and that of $r_i$ as follows
\begin{equation}
    r_j(t_{n+1}) = \alpha_{i,j} r_j(t_n) + \beta_{i,j} r_i(t_n) + \gamma_{i,j} + \eta_{i,j,n}.
    \label{eq:regression_def}
\end{equation}
In the last equation $\alpha_{i,j}$ estimates the self driving force, $\beta_{i,j}$ measures the influence of the return of exchange rate $r_i$, $\gamma_{i,j}$ accounts for possible nonzero mean return of the rate $r_j$, and $\eta_{i,j,n}$ is the noise term. When $\beta_{i,j}$ is zero, then $r_j$ is not caused by $r_i$, otherwise there is causal relationship from returns on rate $i$ to $j$.

%where $\alpha_{i,j}, \beta_{i,j}$ and $\gamma_{i,j}$ are the regression weights while $\eta_{i,j,n}$ is the noise term. The subscript $i$ is used to suggest that the regression is made with the rate $r_i$. In the last expression $\alpha_{i,j}$ estimates the self driving force, $\beta_{i,j}$ measures the influence of the return of exchange rate $r_i$, while $\gamma_{i,j}$ accounts for possible nonzero mean return of the rate $r_j$. When $\beta_{i,j}$ is zero, then $r_j$ is not caused by $r_i$, otherwise there is causal relationship from returns on rate $i$ to $j$. 

The quality of a regression model is usually estimated in terms of the mean of the squared residuals. Concretely, let us denote with $S_0$ the variance of the dependent variable. Correspondingly, let $S$ be the mean of the square of the residuals obtained from regression of the kinds of Eqs. (\ref{eq:self_regression}) or (\ref{eq:regression_def}). The estimate of the quality of the model is then given by the coefficient of determination 
\begin{equation}
    R^2 = 1 - \frac{S}{S_0}.
\end{equation}
When one compares two regression models, nested into each other, one needs a way to compare them. Because the one with more parameters is expected to have better performance than the other, one needs to know whether it significantly outperforms the simpler peer. The respective tool for such estimates is the $F$-test. Consider a simpler regression model with $p_1$ parameters which result in residual sum of squares $S_1$, and its bigger alternative with $p_2$ parameters and residual sum of squares $S_2$ , applied on the same $N$ data items. Then, if both have the same quality in the null hypothesis, the respective $F$-test statistic
\begin{equation}
    F = \frac{\frac{S_1-S_2}{p_2-p_1}}{\frac{S_2}{N-p_2}},
\end{equation}
has $F$ distribution with $(p_2-p_1,N-p_2)$ degrees of freedom. %If one obtains value of $F$ larger than the selected critical value, that corresponds to certain level of confidence, for example 0.01, than the null hypothesis is rejected and the bigger model is accepted as significantly better.

We note that a more general approach can be obtained with the vector autoregression model (VAR)~\cite{lutkepohl2005new}. VAR is a generalization of the previously explained procedure, where the log return $r_j (t_{n+1})$ is regressed with more than one, or even all other returns. Unfortunately, we cannot apply this procedure on the data under study, because many exchange rates have either zero return, or missing values at different moments, and the regression would then be meaningless.

\section{Data}

\subsection{Data source}
%TICK TIME - moment koga ima promena. Tik e najmala promena na cenata

%\textcolor{red}{Our interest is primary to study the whole dataset, instead on focusing on specific time series. So, we have opted to use the minute-based observations of the exchange rates for spotting the possible lagged correlations and causality between certain pairs. The study of tick-by-tick data is much more computationally demanding, so we have used them only in order to determine the possible roots of such lagged correlations.}

%\textcolor{blue}{Iako stoe deka ima, sepak nema brojki za obrtot, pa ne mozhe da procenime koi dvojki povekje se trguvaat}

%\textcolor{red}{Sho da pravime so dupkite vo zapisite. Dali da gi skoknime?}

%\textcolor{blue}{1. Dali da prodolzhime da barame objasnuvanje zoshto se asimetrichni i dali da presmetame LLR? Lead-Lag ratio. Dali mozhe da se dokazhe deka indeksite imaat pogolemo chekanje i povekje raspar megju cenite?}

%\textcolor{red}{Da kazhime zoshto ne rabotime so valuti, ami so kursevi.}

%Empirical study in this work is performed on historical exchange rates obtained from freely available data source \footnote{The data under study was downloaded from www.histdata.com}. 

We use data gathered from \url{www.histdata.com}\footnote{All studied data is made freely available by its publisher.}. The dataset contains highly frequent one-minute exchange rate values only on the bid quotes, which is a slight drawback since generally as price of an asset is considered the mean of the respective bid and ask values. Even though the bid is not simply a drifted version of the mean price, it can serve in the analysis as an estimate for the value from the point of view of the brokers.

%The difference between bid and ask quotes, or the bid-ask spread is also changing and thus the bid is not simply a drifted version of the mean price. However, the bid values can serve for analysis as well.  
%and could be understood as combination of the mean price and the liquidity of the exchange rate that is encoded into the bid-ask spread. Thus, 

We extract exchange rate data for 66 pairs among 33 assets consisting of 19 currencies, 10 indexes of major stock markets, 2 oil types, and gold and silver. The list of assets together with their abbreviations is given in Table~\ref{tab:Abbreviations}. %\The resulting asymmetrical correlation matrix has size $66 \cdot 66$. The positive matrix entries imply that the respective log returns more frequently deviate from the means in the same direction. The negative ones correspond to the opposite -- positive return of one rate is more likely accompanied with negative of the other. The impact is bigger when the absolute value of the respective correlation coefficient is larger. From the results obtained from the real data it was obtained that the mutual influence of the exchange rates is really asymmetric. One can be satisfied with such lagged correlation matrix which tells return on which exchange rate $i$ with certain sign and size will be probably followed by similar return on another rate $j$ one minute later.

%Our empirical studies were performed on subset of freely available foreign exchange historical data obtained from histdata.com. The dataset provides exchange rates for 66 pairs between 33 assets which consist of 19 currencies, 10 indexes of major stock markets, 2 oil types, gold and silver. The respective codes of the assets are provided in the table \ref{tab:Abbreviations}. The database contains  bid and ask quotes of the pairs at every tick, but these are asynchronous because price changes do not occur simultaneously. We have opted to look at the less frequent minute data for which only bid quotes are provided. In studies related with financial time series, as prices of an asset are considered the averages of bid and asset values. Due to the time variation of the bid-ask spread the bid price is not simply drifted representative of such mean. However, the former is an estimation of the worthiness of the exchange rate as seen from the market makers and can serve for analysis too. 

\begin{table}[h!]
  \begin{center}
    \caption{List of studied assets}
   \label{tab:Abbreviations}
    \begin{tabular}{l|l} % <-- Alignments: 1st column left, 2nd middle and 3rd right, with vertical lines in between
    \textbf{Currencies} \\
      \hline      
     AUD -- Australian Dollar & MXN -- Mexican Peso\\
     CAD -- Canadian Dollar & NOK -- Norwegian Krone\\
     CHF -- Swiss Franc & NZD -- New Zealand Dollar\\
     CZK -- Czech Koruna & PLN -- Polish Zloty\\
     DKK -- Danish Krone & SEK -- Swedish Krona\\
     EUR -- EURO & SGD -- Singapore Dollar\\
     GBP -- British Pound & TRY -- Turkish Lira\\
     HKD -- Hong Kong Dollar & USD -- US Dollar\\
     HUF -- Hungarian Forint & ZAR -- South African Rand\\
     JPY Japanese Yen \\
     \hline      
     \textbf{Indexes} \\
     \hline      
     AUX -- ASX 200 & JPX -- Nikkei Index 400 \\
     ETX -- EUROSTOXX 50 & NSX -- NASDAQ 100 \\
     FRX -- CAC 40 & SPX -- S\&P 500 \\
     GRX -- DAX 30 & UDX -- Dollar Index \\
     HKX -- HAN SENG & UKX -- FTSE 100 \\
     \hline      
     \textbf{Commodities} \\
     \hline      
	 BCO -- Brent Crude Oil & XAG -- Silver \\
     WTI -- West Texas Intermediate & XAU -- Gold \\
     \hline
  
    \end{tabular}
  \end{center}
\end{table}

%As is declared at the host of the database, there are gaps present, which means that at some moments data is missing. To overcome such issue we use discard data entries which have missing values. More details about data filtering is provided in the beginning of the following section. The analyses were performed for each of the 36 months in the last three years: 2016, 2017 and 2018. The estimates of the lagged correlations and causality were based on periods of one month because determination of correlations of longer series is computationally more demanding.

\subsection{\label{sec:dataProc}Data preprocessing}

\begin{figure*}[t!]
\begin{adjustwidth}{0in}{0in}
\includegraphics[width=12.7cm]{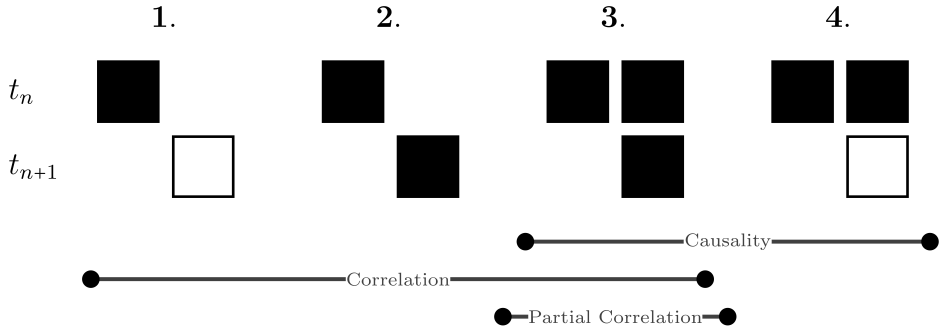}
\caption{Visual representation of the four scenarios of restrictions on the values of the log returns and in which calculations are they used (bottom lines). In all four cases the left square denotes the leader, while the right one(s) correspond to the lagger. Full square denotes that only observations with nonzero value is considered, while empty one means that the log return which is used might have zero value as well. %The four scenarios are as follows: $\mathbf{1}.$ Only leader is always nonzero. $\mathbf{2}.$ Both leader and lagger are nonzero. $\mathbf{3}.$ The leader and the lagger in two consecutive minutes have nonzero returns. $\mathbf{4}.$  The leader and the lagger must be nonzero simultaneously.
\label{fig:case-studies}}
\end{adjustwidth}
\end{figure*}

In spite of the foreign exchange market being highly dynamic, the data which we study includes several cases where a rate has the same value in few consecutive minutes. This means that the respective one-minute returns are zero. In addition, there are gaps with missing data in the time series. We filter out the data from such situations and examine four different lead-lag scenarios, as illustrated in Fig.~\ref{fig:case-studies}.%This situation particularly happens for some market indexes, like the AUX, ETX, FRX, and so on. 
%For that reason in our empirical study on one-minute data, we have paid particular attention to the periods of inactivity.
%Although if at least one of the returns in the correlation pair under study is zero, the product would not increase the sum, it increases the number of summands. 
%\textcolor{blue}{Sinoto bi se zamenilo so crvenoto. These zero returns can be either due to inactivity or due to several interacting effects} \textcolor{red}{Viktor, shto e ova interacting effects?}. \textcolor{blue} {To account for the potential interacting effects we consider four different scenarios for excluding trivial zero returns. The scenarios are illustrated in Fig.~\ref{fig:case-studies}}.
 One can think of the correlations and causal relationships studied in these scenarios as conditional ones, because the calculations are conditioned on the presence of nonzero return on certain variables. In the first scenario, only the leader is restricted to have nonzero return, while the lagger is allowed to be dormant in the following minute. In this aspect, we calculate the lagged correlation only from pairs in which the exchange rate which precedes has only nonzero returns. If in such case one obtains statistically significant correlation, it should mean that, although sometimes the lagger might retain its value in the following minute, when it modifies, the change would likely have the same sign and similar size as the leader had in the past minute. We argue that this is due to market information flowing from the leader towards the lagger. In addition, the statistically significant correlation might represent a \textit{potential statistical arbitrage}, since after observing the return of the leader, fast-profit seekers could take the appropriate position on the lagger. We consider this as potential arbitrage, because even though the lagger's exchange rate would more likely have the right direction of change, the expectation of the profit might not compensate the transaction costs. Obviously, a detailed study involving also data on the transaction costs and ask quotes could reveal whether these lead-lag relationships are indeed profitable. 

The second case corresponds to the correlations obtained from the observations of simultaneous nonzero returns on the leader and the lagger in the next minute. Under such circumstances, one has an estimate on how much of the change of the price of the lagger, which certainly occurred, is possibly influenced by the previous return on the leader. 

In order to calculate the partial correlation, in the third scenario we further specify the lagger to have nonzero return at the same minute as the leader. Such restriction is necessary for removal of the correlation between simultaneous returns of the leader and the lagger and the lagged autocorrelation of the latter, as potential contributors. This scenario was also used for determination of causal relationships.

In the last case we use observations of nonzero returns on the leader and lagger at the same minute, and apply it solely for identifying Granger causality. Under this scenario, one checks for predictability potential based on linear regression when one has two signals -- simultaneous observations of nonzero values of the two relevant returns.

The collected data spans from January 2016 until December 2018. For computational reasons we calculated the quantities of interest for each of the 36 months separately. In representing the rankings, the averages of the monthly results are given.

 Moreover, we point out that even though log returns are usually assumed to be weakly stationary, the empirical data may not exhibit such properties. This might be especially true for foreign exchange markets where the assets display extremely high dynamics and volatility. To statistically test whether the data used here is appropriate for studying lead-lag relationships that we are interested in, we utilize the Dickey-Fuller test~\cite{dickey1979distribution}. This is a standard test used in time series analysis for examining the weak stationarity of a series. Under the null hypothesis, the test assumes that the time series under study has unit root. More precisely, we have applied augmented version of the test \cite{wooldridge2015introductory}, where the difference of consecutive returns $\Delta r_i(t_n) = r_i(t_n) - r_i(t_{n-1})$ is regressed with the past return $r_i(t_{n-1})$, and the previous difference $\Delta r_i(t_{n-1})$. We could not examine lags of the difference of log returns of higher order because the presence of gaps in the data would decrease the number of samples where chains of consecutive differences are available. Correspondingly, this would weaken the statistical power of the test. Also, possible presence of time trend in the series was accounted for. To sum up, we have considered the following regression model
 \begin{equation}
     \Delta r_i(t_n) = \alpha + \delta t_n + \theta r_i(t_{n-1}) + \gamma_1 \Delta r_i(t_{n-1}) + \epsilon_n,
 \end{equation}
where $\alpha$, $\delta$, $\theta$, and $\gamma_1$ are coefficients, and $\epsilon_n$ is the error term. Under the null hypothesis, $\theta=0$ suggests presence of unit root.
 
 The Dickey Fuller test statistic calculated for each month in 2016 separately, and for all exchange rates is reported in the Table \ref{tab:DFTestStat}. Because the critical value of the test statistic for level 0.01 is -3.96 \cite{fuller2009introduction}, one can observe that in each case the null hypothesis of presence of unit root can be rejected with high confidence. Accordingly, one can consider that the time series of log returns are weakly stationary.

 \begin{table}[h!]
\begin{minipage}{\textwidth}
  \begin{center}
    \caption{Dickey-Fuller test  statistic (DFTS) for studied exchange rates in 2016}
    
   \label{tab:DFTestStat}
    \begin{tabular}{|l|r||l|r||l|r|} % <-- Alignments: 1st column left, 2nd middle and 3rd right, with vertical lines in between
          \hline      
    \textbf{Rate} & \textbf{DFTS} & \textbf{Rate} & \textbf{DFTS} & \textbf{Rate} & \textbf{DFTS}\\
      \hline      
     AUD/CAD & -125.6 & EUR/SEK & -113.8 & USD/CAD & -121.8 \\ 
     \hline
AUD/CHF & -122.1 & EUR/TRY & -117.0 & USD/CHF & -122.5 \\ 
     \hline
AUD/JPY & -121.4 & EUR/USD & -131.0 & USD/CZK & -118.8 \\ 
     \hline
AUD/NZD & -126.3 & FRX/EUR & -89.9 & USD/DKK & -122.1 \\ 
     \hline
AUD/USD & -122.0 & GBP/AUD & -123.8 & USD/HKD & -49.2 \\ 
     \hline
AUX/AUD & -79.8 & GBP/CAD & -123.7 & USD/HUF & -111.1 \\ 
     \hline
BCO/USD & -96.4 & GBP/CHF & -123.5 & USD/JPY & -117.4 \\ 
     \hline
CAD/CHF & -122.6 & GBP/JPY & -118.7 & USD/MXN & -113.2 \\ 
     \hline
CAD/JPY & -120.4 & GBP/NZD & -123.2 & USD/NOK & -118.3 \\ 
     \hline
CHF/JPY & -118.5 & GBP/USD & -120.5 & USD/PLN & -116.3 \\ 
     \hline
ETX/EUR & -54.7 & GRX/EUR & -92.5 & USD/SEK & -125.5 \\ 
     \hline
EUR/AUD & -119.0 & HKX/HKD & -83.8 & USD/SGD & -116.4 \\ 
     \hline
EUR/CAD & -122.0 & JPX/JPY & -68.9 & USD/TRY & -103.1 \\ 
     \hline
EUR/CHF & -121.4 & NSX/USD & -105.4 & USD/ZAR & -111.4 \\ 
     \hline
EUR/CZK & -39.6 & NZD/CAD & -123.7 & WTI/USD & -98.4 \\ 
     \hline
EUR/DKK & -28.1 & NZD/CHF & -122.6 & XAG/USD & -108.9 \\ 
     \hline
EUR/GBP & -122.6 & NZD/JPY & -121.1 & XAU/AUD & -117.4 \\ 
     \hline
EUR/HUF & -91.2 & NZD/USD & -121.0 & XAU/CHF & -118.9 \\ 
     \hline
EUR/JPY & -117.7 & SGD/JPY & -118.8 & XAU/EUR & -118.7 \\ 
     \hline
EUR/NOK & -121.7 & SPX/USD & -75.7 & XAU/GBP & -102.1 \\ 
     \hline
EUR/NZD & -122.1 & UDX/USD & -105.9 & XAU/USD & -117.4 \\ 
     \hline
EUR/PLN & -101.8 & UKX/GBP & -89.9 & ZAR/JPY & -110.2 \\ 
     \hline
    \end{tabular}
  \end{center}
  \end{minipage}
\end{table}

 %Before calculating the correlation coefficients and causality relationships we have tested whether the studied time series of log returns are stationary. As verified with the Dickey-Fuller test\cite{dickey1979distribution}, which is the most exploited tool for such purpose the presence of unit root in the series is rejected, which suggests that the returns can be considered as stationary. Tuka Viktor da dopishime referenca, i mozhebi nekoj dopolnitelen komentar.}

Finally, we point that here we have considered exchange rates as assets, instead of using values of currencies or commodities. To obtain the value of a currency one should quote it in terms of another one, taken as base, by using the appropriate exchange rate~\cite{basnarkov2019correlation}. However, one does not always have a direct exchange rate between the appropriate base and any other currency in this market. In such cases two or more exchange rates might be needed for determination of the value of the asset. When many gaps in the data are present, or when some returns are inactive for certain period, getting such values can be problematic. Besides this, it has been found that the choice of basic currency can result in nontrivial difference in the obtained correlations~\cite{keskin2011topology, basnarkov2019correlation}.

\section{\label{sec:results}Results and discussion}

\subsection{Statistical significance of results}

To determine statistical significance of the results, we applied the Bonferroni threshold correction. The Bonferroni correction is standardly used for handling situations when one makes multiple tests simultaneously, as is the case with the overall significance of the correlation matrix~\cite{tumminello2011statistically, curme2015emergence}. For correlations between each of the $N$ assets the Bonferroni correction is $N \cdot N$. Thus, the appropriate threshold for $p$ value at level $0.01$, which is widely applied in the literature~\cite{curme2015emergence}, is $0.01 / (66 \cdot 66)$. This threshold was applied for the two types of lagged correlations and for the Granger causality relationships.

%So, the statistical significance in present study was obtained from the theoretical expression (\ref{eq:P_value_theor}) \textcolor{red}{Mozhe i da ne e Fisherova. Da vidime kaj Tuminelo u Kvant. Fin. vo dodatokot (B5) Kenney, J.F. and Keeping, E.S., Mathematics of Statistics} and the correlation was considered as being non accidental when the respective P value was smaller than $0.01 / (66 \cdot 66)$. Clearly, a reader who is interested in more reliable estimates of the significance of certain correlations would probably need to apply the shuffling approach.

%\begin{figure*}[t!]
%\begin{adjustwidth}{-0.8in}{0in}
%\includegraphics[width=14cm]{PValue.png}
%\caption{Comparison of the P value obtained with theoretical equation (\ref{eq:P_value_theor}) with that resulting from permutation test. \textcolor{red}{Zoshto butstrep?} \label{fig:PValue_compared}}
%\end{adjustwidth}
%\end{figure*}

\subsection{Lagged correlations}

As it was elaborated in Section~\ref{Lagged_corr_theory}, the lagged correlations quantify the relationship between series of one-minute log returns of two exchange rates with one of them trailing the other for a unit period. The existence of statistically significant correlation implies that the one with preceding returns presumably influences the other. We calculated such lagged correlations for all pairs in the dataset and summarized them in the respective $66\times 66$ correlation matrix. Due to the very intensive trading of the currencies, and highly dynamical nature of their values, most of the lagged correlations between exchange rates involving currencies do not pass the statistical significance tests. However, there are many nonzero terms in the lagged correlation matrix, notably those involving stock market indexes.  %For obtaining of that matrix we have applied the Bonferroni correction and the threshold was set to $0.01 / (66 \times 66)$ as in \cite{curme2015emergence}. 
As expected, the correlation matrix is asymmetrical because the influence is not identical in both directions. Out of this lagged correlation matrix we create a directed network where a link from node $j$ to $i$ exists if the rate $i$ has statistically significant correlation with $j$, and where $i$ is the leader. %If both lagged correlations between a pair $(i,j)$ are statistically significant then the link goes out from the node which is weaker leader, which means that the lagged correlation when it is leader is smaller than when it lags. 
%In order to determine which rates are more important leaders we have applied the Page Rank algorithm on the respective weighted directed network. %\textcolor{red}{Tuka za Page Rank. Journal and Country ranking  - SCIMAGO -- \cite{falagas2008comparison}}. As explained in the previous section the direction was chosen to be towards the node which is stronger leader in the pair, while the weight was taken to be the difference of the absolute values of the lagged correlations. If both rates do not mutually influence each other, the weight was taken to be the only lagged correlation present between the pair. 

 In the second, third and fourth columns of Table~\ref{tab:Ranking} we present the ten most influential exchange rates on average, according to the PageRank for the three years under consideration and the three lagged correlation scenarios respectively. The results appear rather surprising, since market indexes with slower dynamics are put on top. This is opposite to previous findings where it was discovered that the more liquid assets lead the others~\cite{kawaller1987temporal, lo1990contrarian, toth2006increasing}.  In particular, the first four entries, all corresponding to such market indexes, appear rather high on the rank list nearly every month. The following ones, such as WTI oil exhibit varying influence in the lagged correlation matrix and are ranked rather differently in different months. The reason that the market indexes are most influential could be the fact that their price dynamics is slower and every change of their value incorporates some market news. %Moreover, the correlations involving the market indexes are not marginal because they pass rather strong significance test that corresponds to the threshold $0.01 / (66 \cdot 66)$.

\begin{sidewaystable}[h!]
  \begin{center}
    \caption{Top ten most influential exchange rates in five cases: LC1, LC2 and LC3 -- lagged correlations in scenarios 1, 2, and 3 respectively; LPC -- most influential rates in lagged partial correlations; C1 and C2 -- ranking of Granger causality significance in scenarios 3 and 4 respectively.}
   \label{tab:Ranking}
    \begin{tabular}{c|c|c|c|c|c|c} % <-- Alignments: 1st column left, 2nd middle and 3rd right, with vertical lines in between
    \textbf{Rank} & \textbf{LC1} & \textbf{LC2} & \textbf{LC3} & \textbf{LPC} & \textbf{C1} & \textbf{C2}\\
      \hline      
     1 & ETX/EUR & ETX/EUR & ETX/EUR & ETX/EUR & ETX/EUR & ETX/EUR\\
     2 & JPX/JPY & JPX/JPY & JPX/JPY & JPX/JPY & JPX/JPY & JPX/JPY\\
     3 & AUX/AUD & AUX/AUD & AUX/AUD & SPX/USD & SPX/USD & SPX/USD\\
     4 & SPX/USD & SPX/USD & SPX/USD & AUX/AUD & AUX/AUD & AUX/AUD\\
     5 & USD/CZK & USD/CZK & USD/CZK & WTI/USD & USD/CZK & USD/CZK\\
     6 & WTI/USD & WTI/USD & WTI/USD & USD/CZK & WTI/USD & WTI/USD\\
     7 & EUR/USD & USD/HUF & USD/HUF & NZD/USD & AUD/USD & AUD/USD\\
     8 & USD/HUF & ZAR/JPY & ZAR/JPY & AUD/USD & NZD/USD & UDX/USD\\
     9 & ZAR/JPY & EUR/USD & AUD/USD & USD/DKK & UDX/USD & NZD/USD\\
     10 & UDX/USD & UDX/USD & EUR/USD & BCO/USD & BCO/USD & USD/DKK\\
     
      \hline
    \end{tabular}
  \end{center}
\end{sidewaystable}

To validate whether some correlations are not a random result, but simply correspond to low influence, we also look at the results without the Bonferroni correction. The results for the statistically significant lagged correlations, which pass the threshold $p=0.01$ in each of the studied 36 months are displayed in Table~\ref{tab:PersistentCorr}. These results correspond to the first three scenarios. One can note that mostly the same correlations are present in the three situations. Only in few cases some additional correlation pairs appear when either the return in the leader is nonzero, or both the leader and the lagger in the next minute have nonzero returns. Although accidental emergence of correlation among $N$ uncorrelated time series with such significance threshold would not be a surprise, its appearance in 36 months consecutively suggests that some relationships between all such pairs really exist. From the same table one can notice some interesting patterns. First, dominant leaders are the same four stock market indexes AUX, ETX, JPX, and SPX, that appear as top sources of information spillover as detected with the PageRank. They influence the other indexes FRX, GRX, HKX, NSX, UKX and UDX. It is interesting that the other US-based index NSX is not as influential as the SPX. In addition, one can note that as laggers appear many exchange rates involving the Japanese Yen. When the above mentioned leading market indexes gain value, the Yen seems to react by weakening with respect to other assets. Another intriguing observation is the lowering of the value of gold ounce, XAU, quoted in AUD, when these indexes rise. One can also note that rise or fall of the MXN with respect to the US dollar, is followed by similar behavior of the TRY. These two minor currencies have been also found to be mutually related in a previous study \cite{basnarkov2019correlation}. Finally, it is interesting that the fall of the MXN with respect to the USD, is followed by similar behavior of the ZAR with respect to the JPY. %\textcolor{blue}{Da napravime grafik za ovaa tabela i da vidime kolku kje bide interesen.}

%\textcolor{red}{Ova da go proverime: We should mention that there are much more pairs of exchange rates which are not statistically significantly correlated only in few of the 36 considered months. As such examples, the rises of the prices of two oil types WTI and BCO, incur similar behavior in near future of the prices of the CAD and NOK with respect to the other currencies. One could easily understand this because among the countries which have currencies present in the data under study, Canada and Norway are the highest ranked in the list of oil exporters} \textcolor{red}{referenca}. Some readers might find similar examples about currencies of their interest. 

\begin{table}[h!]
\begin{minipage}{\textwidth}
  \begin{center}
    \caption{Persistent statistically significant correlations.}
   \label{tab:PersistentCorr}
    \begin{tabular}{l|l} % <-- Alignments: 1st column left, 2nd middle and 3rd right, with vertical lines in between
    \textbf{Leader} & \textbf{Positively correlated lagger} \\
      \hline      
     AUX/AUD & AUD/JPY, CAD/CHF, CAD/JPY, FRX/EUR, \\ & GRX/EUR, HKX/HKD, NSX/USD, NDZ/JPY, \\ & SGD/JPY, SPX/USD\footnote{\label{ftnt:L_or_LF}Persistent correlation when either the leader, or both the leader and the lagger have nonzero returns (scenarios 1 and 2).}, UDX/USD, UKX/GBP,\\ & USD/JPY\\
     \hline    
     ETX/EUR & AUD/JPY, CAD/JPY, FRX/EUR, GBP/JPY, \\ & GRX/EUR, HKX/HKD, NSX/USD, NDZ/JPY,\\ 
     & SGD/JPY, SPX/USD, UDX/USD, UKX/GBP,\\ & USD/JPY\\
     \hline    
     JPX/JPY & AUD/JPY, CAD/CHF, CAD/JPY, CHF/JPY\footref{ftnt:L_or_LF}, \\ & 
     EUR/JPY, FRX/EUR, GBP/JPY, GRX/EUR,\\ & HKX/HKD, NSX/USD, NDZ/JPY, SGD/JPY, \\ & SPX/USD\footref{ftnt:L_or_LF}, UDX/USD, UKX/GBP, USD/JPY\\
     \hline    
     SPX/USD & AUD/JPY, CAD/JPY, SGD/JPY\footref{ftnt:L_or_LF}\\
     \hline
     USD/MXN & USD/TRY\\
     \hline
\textbf{Leader} & \textbf{Negatively correlated lagger}\\
      \hline
     AUD/USD & USD/TRY \footnote{Persistent correlation when only the leader has nonzero return (scenario 1). \label{ftnt:L}}\\
      \hline
     ETX/EUR & XAU/AUD\\
     \hline
     JPX/JPY & XAU/AUD\\
     \hline
     SPX/USD & XAU/AUD\\
     \hline
     USD/MXN & ZAR/JPY\footref{ftnt:L}\\
     \hline
    \end{tabular}
  \end{center}
  \end{minipage}
\end{table}

%The lagged correlation coefficients within this market are not very large, as one can expect due to the efficient market hypothesis, but they are not always very small either. In the figure \ref{fig:Correlation_histogram} is shown the histogram of the statistically significant lagged correlation coefficients for all months in 2017 \textcolor{red}{Dali ni treba ova?}. One can see that although rarely, noticeable values appear as well. Because the vertical axis is in logarithmic scale, the histogram suggests that the correlation coefficients have exponential, possibly Gaussian distribution. Whether this is correct, and if there is mechanism which could explain this is an open issue.

%\begin{figure*}[t!]
%\begin{adjustwidth}{-0.8in}{0in}
%\includegraphics[width=14cm]{CorrHist.png}
%\caption{Histogram of the statistically significant lagged correlation coefficients for 2017. \label{fig:Correlation_histogram}}
%\end{adjustwidth}
%\end{figure*}

\subsection{Lagged partial correlations}

%The partial correlation in general is quantitative measure of the linear statistical dependence of a pair of random variables, when certain number of external variables are considered as known. In our case, we wanted to remove the influence of the serial autocorrelation of the log return, and the contemporaneous correlation between two exchange rates as possible causes of lagged correlation. Because lagged partial correlation can emerge by chance as well, those which are not statistically significant with the threshold $0.01/(66\cdot 66)$, were filtered out.
We quantify the importance of the exchange rates as determined from the lagged partial correlations through the PageRank algorithm, as well. In the fifth column of Table~\ref{tab:Ranking}, we show the ranking of the most influential exchange rates. The same four major market indexes, as in the lagged correlation case, appear at the top of the rankings. When applying the ordinary critical value of $0.01$ instead of the Bonferroni correction
%,In order to see whether some weaker lagged partial correlations exist, but are removed by applying the Bonferroni correction, we have applied the ordinary threshold $0.01$ as well. Again, we have considered as non accidental only those that are statistically significant in all 36 months.
we find similar results as in the previous section. These results are provided in Table~\ref{tab:PersPartCorrAndCaus}.
%Such lagged partial correlation pairs are provided in the table \ref{tab:PersPartCorrAndCaus}. First, one will notice similar relationships as those obtained with the ordinary lagged correlations.
However, these results also suggest why at the top is the European stock market index ETX.
%When considering the most important exchange rates, it comes apparent why at the top is the European stock market index ETX. 
Namely, the log returns on it, lead the returns of all market indexes AUX, FRX, GRX, HKX, JPX, NSX, SPX, and UKX. It is influenced by the FRX and GRX, which is a feedback relationship, something that was not present in the ordinary correlations. This influence is not symmetrical and is more pronounced from the ETX towards FRX and GRX, than in the opposite direction.

We find that other bidirectional partial correlations are those between the two oil brands, and between TRY and ZAR when quoted in US dollars. We remark that in the ordinary case, these currencies did not possess any mutual lagged correlation. Moreover, our results show that the value of TRY and ZAR in US dollars seems to be also similar in their trailing of the dynamics of the two Oceanic currencies AUD and NZD as quoted in the USD. The returns on AUD/USD also show sign of having some influence on those on SGD/USD. Particularly important is the fact that the return on the silver in USD, precedes similar behavior of the gold in the next minute. In a related observation involving precious metals, it appears that the CHF adjusts its value in terms of gold, after the neighboring EUR has done it one minute before. When any of the MXN, TRY, or ZAR gains its value in USD,  ZAR is induced to behave similarly with respect ot the JPY. Also, it is likely that the MXN has some influence on the ZAR which was not observed from the ordinary lagged correlations. Finally, interesting observations are those involving currencies from the Eurozone. The value of PLN in USD seems to influence the HUF quoted in the same currency base. It is also peculiar that the value of the minor currency CZK given in terms of US dollars, influences the major EUR, but not the opposite.

%Finally, differently from the results discovered in the simple lagged correlation analysis, %ZAR quoted in USD leads MXN to do so when considered ordinary lagged correlation, but not the opposite, in the partial correlation case the roles are reversed --

\subsection{Granger causality}

%The presence of Granger causality was determined by using linear regression of the future value of log returns on certain exchange rate $r_j(t_{n+1})$, with its preceding one $r_j(t_n)$ and with preceding return on another rate $r_i(t_n)$. When adding the external rate $r_i(t_n)$ improves the predictability of $r_j(t_{n+1})$, by decreasing the regression residual, one concludes that the exchange rate $i$ Granger causes the rate $j$.
The tests for presence of Granger causality were performed under scenarios 3 and 4, where both the leader and lagger must have nonzero returns simultaneously, while in scenario 3 additionally the lagger also needs to have non-vanishing returns in two consecutive periods. %The threshold for statistical significance was set to such $F$ value that corresponds to $p$ value equal to $0.01/(66\cdot 66)$. 
The statistically significant causal relationships were presented in a directed network and the importance of exchange rates was again determined with the PageRank algorithm. We note that the networks which we have considered for depicting the causal relationships are not weighted, which means that a link between two pairs exists only if the source node is influenced by the target node, and the weight of each such link is one. As expected, the rates involving market indexes appear at the top of the list. The rankings of ten rates with strongest causal influence are provided in the last two columns in table~\ref{tab:Ranking}, where similar ordering with the other two measures (lagged correlations and lagged partial correlations) is observed. Clearly, it can be argued that the lagged correlations, both partial and ordinary, suggest that a causal relationship is present between a leader and its lagger. By lowering the threshold for statistical significance to $0.01$, as in previous cases, we obtain almost the same pairs of exchange rates as in the case of partial correlations. The only exception is the existence of statistically significant partial lagged correlation between the JPX/JPY as leader and the CHF/JPY. The causality relationships under both scenarios between the same pair does not pass the threshold for only one month, December in 2017. This might be result of statistical nature and one could speculate that both scenarios produce the same results.

We state that one might be suspicious in the predictability under scenario 3, where we determine causality when the lagger is constrained to have nonzero return in the next minute. However, in the weaker case, when non-vanishing returns of the leader and the lagger are observed simultaneously, one can use them as signals, and make a regression on the future value of the return on the lagger. %Nevertheless, it might not be profitable at all, when one considers the bid and ask prices, or the transaction costs, which we unfortunately cannot verify from the data under study.

\begin{table}[h!]
  \begin{center}
    \caption{Persistent statistically significant partial correlations and causality relationships.}
   \label{tab:PersPartCorrAndCaus}
    \begin{tabular}{l|l} % <-- Alignments: 1st column left, 2nd middle and 3rd right, with vertical lines in between
    \textbf{Leader} & \textbf{Same direction} \\
      \hline      
     AUX/AUD & AUD/JPY, CAD/CHF, CAD/JPY, FRX/EUR, \\ & GBP/JPY, GRX/EUR, HKX/HKD, NSX/USD, \\ & NDZ/JPY, SGD/JPY, SPX/USD, UDX/USD, \\ & UKX/GBP, USD/JPY, ZAR/JPY\\
     \hline    
     BCO/USD & WTI/USD \\
     \hline
     ETX/EUR & AUD/JPY, AUX/AUD, CAD/CHF, CAD/JPY, \\ & FRX/EUR, GBP/JPY, GRX/EUR, HKX/HKD, \\ & JPX/JPY, NSX/USD, NDZ/JPY, SGD/JPY, \\ & SPX/USD, UDX/USD, UKX/GBP, USD/JPY\\
     \hline 
     FRX/EUR & ETX/EUR, GRX/EUR \\
     \hline
     GRX/EUR & ETX/EUR\\
     \hline
     JPX/JPY & AUD/JPY, CAD/CHF, CAD/JPY, CHF/JPY \footnote{Only the partial lagged correlation is persistent in the considered 36 months.\label{ftnt:Part_Corr}}, \\ & EUR/JPY, FRX/EUR, GBP/JPY, GRX/EUR,\\ & HKX/HKD, NSX/USD, NDZ/JPY, SGD/JPY, \\ & SPX/USD, UDX/USD, UKX/GBP, USD/JPY\\
     \hline    
     SPX/USD & AUD/JPY, CAD/JPY, GRX/EUR, NSX/USD, \\ & NZD/JPY, SGD/JPY, UDX/USD\\
     \hline
     USD/CAD & USD/ZAR \\
     \hline
     USD/MXN & EUR/TRY, USD/TRY, USD/ZAR\\
     \hline
     USD/PLN & USD/HUF \\
     \hline
     USD/TRY & USD/ZAR \\
     \hline
     USD/ZAR & EUR/TRY, USD/TRY\\
     \hline
     WTI/USD & BCO/USD \\
     \hline
     XAG/USD & XAU/USD \\
     \hline
     XAU/EUR & XAU/CHF \\
     \hline
     ZAR/JPY & SGD/JPY \\
     \hline
\textbf{Leader} & \textbf{Opposite direction}\\
      \hline
     AUD/USD & USD/SGD, USD/TRY, USD/ZAR\\
     \hline
     AUX/AUD & XAU/AUD\\
      \hline
     ETX/EUR & XAU/AUD\\
     \hline
     JPX/JPY & XAU/AUD\\
     \hline
     NZD/USD & USD/TRY, USD/ZAR\\
     \hline
     SPX/USD & XAU/AUD\\
     \hline
     USD/CZK & EUR/USD \\
     \hline
     USD/MXN & ZAR/JPY\\
     \hline
     USD/TRY & ZAR/JPY\\
     \hline
     USD/ZAR & ZAR/JPY\\
     \hline
     ZAR/JPY & EUR/TRY\\
     \hline
    \end{tabular}
  \end{center}
\end{table}

Since the foreign exchange is known to be very liquid, one can not expect significant lags that span longer than one minute. To empirically check this hypothesis we also made an analysis where we examine whether these lagged relationships persist for longer periods. For this purpose we re-estimated our metrics for lead-lag relationships by taking into account for lagger returns that trail two minutes. We found that there are some pairs of exchange rates that pass the significance threshold of 0.01. However, they are not consistent. In particular, no such pair passed the test for all months in 2016. This is a further indication that the relevant lagging time scale is of order of one minute. Due to spatial constraints, we do not show these results, but make them available upon request.

\section{\label{sec:conclusions}Conclusions}

%\textcolor{red}{TO provide the readers with a feeling of the size of the correlation we have put in table XXX the largest correlation coefficients.}
To conclude, here we studied the relations between pairs of exchange rates where one exchange rate lags the other for one minute, and considered three different approaches. The first one considers lagged correlations which quantify similarity of log returns between lead and lagged exchange rate. The second approach restricts the lagged correlations to partial, in order to account for the potential autocorrelation. Finally, the last approach tries to uncover the causal relationships in the foreign exchange market in the sense of Granger. 

The existence of statistically significant lagged correlations shows that, even though, the foreign exchange market is known to have very fast dynamics, information spreading is not instantaneous. We discovered that the rates which cause others to follow their dynamics are mostly those which involve stock market indexes. Observing changes in the value of an index, implies that certain currencies, or market indexes would more likely gain, while others would loose value. This was further confirmed by the calculation of the lagged partial correlation between the leader and the lagger exchange rate. By applying the PageRank algorithm on the statistically significant correlations in both cases we found that the most influential rates are indeed those involving market indexes. The same conclusions hold even when we estimated the Granger causality between pairs of exchange rates. When lagging period is two minutes, these lead-lag effects disappear, which suggests that the typical lagging time is of the order of one minute.
%Using past return on another exchange rate does really improve the linear regression model of some rate, than using its past observation only. Again, such prediction improvement is noticed when one applies rates between certain market index and its domestic currency.

In spite of the fact that the existence of lagged correlations and causality challenges the efficient market hypothesis in foreign exchange markets, we are still uncertain of the existence of arbitrage. In order for this phenomenon to be proven, one should verify whether the potential statistical-arbitrage-based profit 
can overcome the transaction costs. Investigating the presence of arbitrage should be a topic of future research which, besides the bid quotes, requires additional information such as ask price, transaction costs and time needed to complete the transaction.

%Within the database that we had considered many currencies were not present.

%As a final remark, we state that the presented study should be seen more as a concept of analysis rather than a complete picture of the influence trajectories in the foreign exchange market. 

As a final remark, we state that here we did not investigate the temporal dynamics of the discovered correlation patterns. A more detailed study which captures the dependencies among these temporal correlations could reveal novel insights for the nature of the lead-lag relationships in foreign exchange markets.

%reveal whether these relationships strengthen or weaken in certain periods of the day or week, or even completely disappear. %A such analysis, should be coupled with Analysis of the highest frequency tick-by-tick data which keeps track of every change of the value of the exchange rates could help to understand or bring novel knowledge on these correlation and causality relationships.

\section{Acknowledgement}

We are grateful to \url{histdata.com} for freely sharing their data. This research was partially supported by the Faculty of Computer Science and Engineering at ``SS. Cyril and Methodius'' University in Skopje, Macedonia and by DFG through grant ``Random search processes, L\'evy flights, and random walks on complex networks''.

%\bibliographystyle{unsrt}
%\bibliography{LaggedCN}
\end{document}